# TLEP, first step in a long-term vision for HEP

M. Koratzinos, A.P. Blondel, U. Geneva, Switzerland; R. Aleksan, CEA/Saclay, France;
P. Janot, F. Zimmermann, CERN, Geneva, Switzerland; J. R. Ellis, King's College, London;
M. Zanetti, MIT, Cambridge, USA.

*input given to M. Klute, co-convenor, group II: Frontier Capabilities; Snowmass process 2013.*

**Introduction**

The discovery of H(126) [1] [2] is a triumph of the Standard Model (SM). The mass of this Higgs boson is consistent with expectations from precision measurements assuming no new physics, but also with those of a number of extensions of the SM invoked to solve the hierarchy problem. It is also consistent with straightforward extrapolation to the Planck scale, so that the need for new physics at the TeV scale is a totally open question.

On a more practical side, the H(126) mass is small enough to allow detailed studies to be carried out with an $e^+e^-$ collider operating near the ZH production maximum, 240GeV in the centre-of-mass ($E_{CM}$), provided a large enough luminosity can be obtained. In order to test the existence of Physics at the TeV scale, a precision of a few per mil on the H(126) couplings to bosons and fermions is called for. Another way to test the existence of new physics beyond what could be observed directly at the LHC, is to sharpen considerably the precision tests of the Electroweak Theory, i.e. the W and Z masses, the top quark mass, and Z peak observables such as the Z width, the polarization and charge asymmetries, the b partial width, etc.

For the study of the Higgs boson, the type of $e^+e^-$ collider, linear or circular, is of little importance, what matters are the deliverable luminosity, maturity of design, risk, timescale and cost. For Electroweak precision measurements (such as $m_Z$, $\Gamma_Z$, $m_W$), the availability of precise energy calibration and of well known longitudinal polarization are also essential. TLEP offers a very competitive proposition compared to linear colliders of similar timescale and cost as it benefits from three unique characteristics of circular machines: i) high luminosity and reliability, ii) the availability of several interaction points, iii) superior beam energy accuracy.

TLEP [3] [4] is an $e^+e^-$ storage ring of 80-km circumference that can operate with very high luminosity from the Z peak (90 GeV) to the top quark pair threshold (350 GeV). It can achieve transverse beam polarization at the Z peak and WW threshold, giving it unparalleled accuracy ($\leq$100 keV per measurement) on the beam energy calibration by resonant depolarization. A preliminary study indicates that an 80 km tunnel could be constructed around CERN [5]. Such a tunnel would allow a 100 TeV proton-proton collider to be established in the same ring (VHE-LHC), offering a long term vision.

For a given RF power, the luminosity of a storage ring collider rises linearly with its circumference. For a given tunnel size, and assuming the machine operates at the beam-beam limit, luminosity rises linearly with the total dissipated synchrotron radiation (SR) power (which is proportional to the total RF power available). Therefore, the analysis can be scaled to different machine circumferences, including an existing tunnel like that of LEP/LHC [6]. The energy reach depends strongly on the machine circumference (a machine of 27-km circumference can reach 240 GeV, the limit is over 350 GeV for a machine of 80km circumference).

**The accelerator**

TLEP is a storage ring with superconducting RF and low $\beta^*$ insertions, operating at fixed field and fed by an accelerator situated in the same tunnel for continuous top-up injection. Multi-bunch operations are necessary for high luminosity below the top energy, thus separated beam pipes should be foreseen for $e^+$ and $e^-$ beams. A first version of the parameters of TLEP has been produced (see Table 1 for the main parameters) [3]. The SR power dissipated in the tunnel is a design parameter which has been fixed to 100MW. In order to achieve high luminosities, $\beta^*_y$ has been fixed to 1mm, which, compared to the longitudinal size of the beams (2-3mm), gives an hourglass factor of around 0.7. For TLEP-H ($E_{CM}$= 240 GeV) and TLEP-t (350 GeV), beamstrahlung reduces the beam lifetime significantly [7]. For successful operation the beam lifetime has to be longer than the refilling time. The Beamstrahlung lifetime depends on the momentum acceptance, on the number of electrons in a bunch and on the horizontal and longitudinal beam sizes (but not on the vertical beam size). We have simulated the beamstrahlung effect using a detailed collision simulator, Guinea-pig [8]. The simulations give reasonable lifetimes for minimal values momentum acceptance of 2.5% and an emittance ratio $\kappa_\epsilon \equiv \epsilon_x / \epsilon_y$ of 500. These parameters require careful design and procedures but are achievable. Both the optics design to ensure such large momentum acceptance and the design of alignment procedures and online corrections needed to ensure the small emittance ratio will be key elements of the accelerator design. A high horizontal to vertical emittance ratio of 1000 is routinely achieved at Synchrotron radiation facilities and should be achievable in TLEP with modern beam instrumentation and perhaps active magnet supports.

In Higgs factory mode a luminosity of $5\times10^{34}$/cm$^2$/s is achieved for each of four IPs for $E_{CM}$ = 240 GeV. We have considered a storage ring with four IPs to allow straightforward extrapolation from LEP2 [9], which gives beam-beam parameter values around 0.1 per IP. If it is decided to have fewer interaction regions, the total facility luminosity is expected to scale roughly as $\sqrt{N_{IP}}$. Most of the components of the proposed superconducting RF system are readily available for frequencies around 700-800 MHz. A gradient of 20MV/m requires 600m of acceleration and a total of 900m of RF cavities, giving an RF system size similar to that of LEP2. Higher accelerating gradients (35MV/m) are achievable today but lead to excessive power requirements for CW operation, which are unnecessary for this project. The design study will address the optimization of the RF system as well as a dedicated R&D to increase the power efficiency of the system beyond the already high value of 50-60% achievable today in CW mode.

A unique capability of circular machines is the beam energy measurement accuracy using resonant depolarization, which allows an instantaneous precision of better than 100 KeV on the beam energy. The beam energy spread being smaller at TLEP than at LEP, transverse beam polarization should be available from the Z pole up to at least 80 GeV per beam. Running with a few dedicated non-colliding bunches will allow the energy to be measured continuously, allowing measurements of the Z mass and width with a precision of 0.1 MeV or better and the W mass with a precision of 1 MeV or better. In addition, movable spin rotators as designed for HERA would allow a program of longitudinal polarized beams at the Z peak, resulting, for one year of data taking, in a measurement of the beam polarization asymmetry with a precision of the order of $10^{-5}$ – or a precision on $sin^2\theta_W^{eff}$ of the order of $10^{-6}$.

**Table 1:** Key accelerator parameters for TLEP operating at the Z pole, the WW threshold, the HZ maximum and the ttbar threshold

|  | TLEP Z | TLEP W | TLEP H | TLEP t |
|---|---|---|---|---|
| $E_{beam}$ [GeV] | 45 | 80 | 120 | 175 |
| circumf. [km] | 80 | 80 | 80 | 80 |
| beam current [mA] | 1180 | 124 | 24.3 | 5.4 |
| #bunches/beam | 4400 | 600 | 80 | 12 |
| #$e-$/beam [$10^{12}$] | 1960 | 200 | 40.8 | 9.0 |
| horiz. emit. [nm] | 30.8 | 9.4 | 9.4 | 10 |
| vert. emit. [nm] | 0.07 | 0.02 | 0.02 | 0.01 |
| bending rad. [km] | 9.0 | 9.0 | 9.0 | 9.0 |
| $\kappa_\varepsilon$ | 440 | 470 | 470 | 1000 |
| mom. c. $\alpha_c$ [$10^{-5}$] | 9.0 | 2.0 | 1.0 | 1.0 |
| $P_{loss,SR}$/beam [MW] | 50 | 50 | 50 | 50 |
| $\beta^*_x$ [m] | 0.5 | 0.5 | 0.5 | 1 |
| $\beta^*_y$ [cm] | 0.1 | 0.1 | 0.1 | 0.1 |
| $\sigma^*_x$ [$\mu$m] | 124 | 78 | 68 | 100 |
| $\sigma^*_y$ [$\mu$m] | 0.27 | 0.14 | 0.14 | 0.10 |
| hourglass $F_{hg}$ | 0.71 | 0.75 | 0.75 | 0.65 |
| $E^{SR}_{loss}$/turn [GeV] | 0.04 | 0.4 | 2.0 | 9.2 |
| $V_{RF}$,tot [GV] | 2 | 2 | 6 | 12 |
| $\eta_{max,RF}$ [%] | 4.0 | 5.5 | 9.4 | 4.9 |
| $\xi_x$/IP | 0.07 | 0.10 | 0.10 | 0.10 |
| $\xi_y$/IP | 0.07 | 0.10 | 0.10 | 0.10 |
| $f_s$ [kHz] | 1.29 | 0.45 | 0.44 | 0.43 |
| $E_{acc}$ [MV/m] | 3 | 3 | 10 | 20 |
| eff. RF length [m] | 600 | 600 | 600 | 600 |
| $f_{RF}$ [MHz] | 700 | 700 | 700 | 700 |
| $\delta^{SR}_{rms}$ [%] | 0.06 | 0.10 | 0.15 | 0.22 |
| $\sigma^{SR}_{z,rms}$ [cm] | 0.19 | 0.22 | 0.17 | 0.25 |
| $\mathcal{L}$/IP[$10^{32}$cm$^{-2}$s$^{-1}$] | 5600 | 1600 | 480 | 130 |
| number of IPs | 4 | 4 | 4 | 4 |
| beam lifet. [min] | 67 | 25 | 16 | 20 |

**Technical challenges**

The **efficiency of the RF system**, a main power consumption and cost driver, will be the subject of the main dedicated hardware R&D. While the efficiency for CW operation is much higher (50-60%) than for pulsed operation of the linear colliders, there remain substantial gains to make. Strong synergies exist with nearly CW machines such as the ESS and other multi-MW systems operating around 800 MHz.

**Optics with low β\* values and large momentum acceptance:** to fight beamstrahlung requires a momentum acceptance in the range 2-2.5%. This requires a dedicated design effort followed by extensive beam tracking in presence of beam-beam effects.

**Procedures to achieve low vertical emittance:** a high horizontal to vertical emittance ratio is a good mitigation technique against beamstrahlung beam lifetime limits. LEP achieved an emittance ratio of 250, but modern light sources achieve values higher than 2000 [10]. The challenge is to design the procedures which, combining beam instrumentation and machine alignment, would allow a low vertical emittance, approaching the values achieved at light sources, to be achieved reproducibly.

**Top-up injection:** For the targeted 1000 s overall beam lifetime, 1% of the beam needs to be replenished every 10 s. This calls for an injector ring, continuously ramping and injecting in the collider ring. In the SPS used in electron injection mode the ramping speed was in excess of 60 GeV/s. The details and integration of the injector require a dedicated design.

**An integrated magnet and beam pipe design** TLEP operation at the Z, WW and ZH requires many bunches (4400, 600 and 80 in the current design), therefore a separate beam pipe for $e^+$ and $e^-$ is called for together with a twin or double magnet design. To simultaneously address vacuum, heat extraction and other issues without impeding magnet operation is a complex design problem.

**Integration issues** with the future VHE-LHC proton collider should be addressed early on and possible synergies identified. The two projects should enjoy very close collaboration on the technical side.

**Polarization**: Much can to be learned from LEP and HERA. Two transverse polarimeters need to be designed for $e^+$ and $e^-$ beams. The 'sterile' bunch operation needed for continuous calibration needs special care. Spin rotator design longitudinal polarization operation at the Z peak and selective bunch depolarization are required. Wigglers are necessary at the Z pole and induce large local power deposition. Achieving longitudinal spin states at higher energies requires a clever arrangement of the experimental spin rotators and is not guaranteed at this point.

**Power consumption**

The luminosity yield of TLEP is proportional to the SR power dissipated in the ring, which is proportional to the RF power. The TLEP design has been performed assuming 100MW of power dissipation in the tunnel (around 1kW per meter of bend), which defines a total power consumption of almost 300 MW in the present state-of-the-art.

We estimate the total RF system efficiency to be 54%-59% [3]. A thyristor 6-pulse power converter for this application has an efficiency of 95%, whereas a switch mode converter runs at 90% efficiency. A klystron run at saturation (as in LEP2) without headroom for RF feedback runs at a 65% efficiency. We consider that fast RF feedback is not necessary for TLEP. RF distribution losses are 5-7%. To estimate the cryogenic power consumption, we use the LHC figures (900W/W at 1.9 K) to arrive at 23 MW at 175 GeV (fundamental frequency dynamic load only). We estimate that the final power consumption would be 1.5 times the dynamic load consumption (to account for static heat loads, HOM dissipation in cavities, overhead for cryogenics distribution etc.), leading to a consumption of 34 MW at 175 GeV. The RF power budget of the accelerator ring is included in this calculation, as the total current in both rings is constant, with the exception of the ramp acceleration power: for a 1.6 s ramp length and 155 GeV energy swing, the total ramp power is estimated to be 5 MW. The power requirements of the RF system at 120 and 175 GeV are summarised in Table 2. For lower energies the power will not exceed the values quoted here. As mentioned above a dedicated RF power efficiency R&D will have very large pay-off in terms of construction and operating costs.

**Table 2:** Preliminary RF power consumption within 2013 state-of-the-art.

|  | TLEP 120 | TLEP 175 |
|---|---|---|
| RF systems | 173-185 MW | |
| cryogenics | 10 MW | 34 MW |
| top-up ring | 3 MW | 5 MW |
| Total RF | 186-198 MW | 212-224 MW |

The power consumption of the rest of the systems is discussed in [3] and adds another 80 MW of power (excluding the experiments). Table 3 shows the breakdown of power consumption at 175 GeV. Consumption at different energies will not exceed this number. We are in contact with the energy optimization group at CERN [11] to find the best ways to utilize the heat generated during operation.

**Table 3:** Preliminary TLEP power consumption at 175 GeV, within 2013 state-of-the-art.

| *Power consumption* | TLEP 175 |
|---|---|
| RF including cryogenics | 224MW |
| cooling | 5MW |
| ventilation | 21MW |
| magnet systems | 14MW |
| general services | 20MW |
| **Total** | ~280MW |

**Main cost drivers**

TLEP is a project at its infancy, therefore a detailed cost estimate does not exist yet. However, the main cost drivers, where effort will be put to reduce costs, have been identified.

It should be highlighted that TLEP is envisaged to be build next to an existing large laboratory, CERN. The existence of a large, mature laboratory next to TLEP historically has helped keeping costs low.

The most expensive ingredient of TLEP is the tunnel and its infrastructure. This, however, should be seen as an investment for CERN since the tunnel could later house next suite of hadron/e-p/ion collider(s) including a very high energy proton collider (VHE-LHC), ensuring more than 50 years of top-tier research. Current estimates for the cost of the tunnel are in the range 2.5-4.5BCHF. These are estimates assuming the CERN site (European accounting throughout).

For the TLEP machine proper the main cost driver is the RF system with its cryogenic infrastructure. If one was to build TLEP now with existing technology and costs, the total cost of the RF system is estimated to be around 1BCHF, out of which two-thirds goes for RF power (converters and klystrons) and one-third to RF cavities and cryomodules [12]. We have here used the SPL item cost.

A cost estimate for the rest of the project does not exist yet, but extrapolations from LEP are possible. The total cost of the TLEP machine proper without the civil engineering is estimated to be about 2-3BCHF with large error bars for the moment, the largest item being the cost of the RF system as mentioned above.

**Design study**

A design study to explore the capabilities and challenges of TLEP has just started [13] and an interim structure with a steering group and an international advisory committee has been defined. Three main areas of work have been identified, accelerator, detector and phenomenology studies, each with an appointed coordinator. More than 200 collaborators have signed up to contribute to the study.

**Timescale**

The aim of the study is to produce a conceptual design report by 2015 and a more detailed technical document by 2018, by which time the first results of the nominal energy run of the LHC would be available. These results would be crucial for defining strategy for High Energy Physics for the next 20-30 years, and TLEP will be ready with a complete report to aid in the process.

Tunnel construction can commence while the LHC is still running, shortly after an eventual approval of the project around 2020. Aim is for first physics at around 2030.